# A simplified approach to the improved semiclassical method

**N.N.Trunov**

trunov@vniim.ru

*D.I.Mendeleyev Institute for Metrology*

*Russia, St.Peterburg. 190005 Moskovsky pr. 19*

(Dated: August, 19, 2013)

**Abstract**. The exact semiclassical quantization condition represents a cumbersome series expansion, so that only the main term of it is usually taken into account. We propose a way to find next terms without new additional calculations. Results are exact for a wide class of actual potentials.

## 1. Introduction

The semiclassical approximation remains valid as the simplest and most universal method for solving problems where the exact solutions are cumbersome or unknown – not only in the quantum theory but also in various fields of science. Advantages of this method as compared with the computer's numerical ones are not only its simplicity but also a clear physical meaning of each stage of calculation. For example, an effective quantum number for centrally symmetric problem was constructed [1] and used for the Periodic system of elements and for nanosystems [2].

The full semiclassical quantisation condition includes a cumbersome series expansion. That is why only the main approximation is mostly used. Several attempts have been made in order to achieve more simple expressions taking into account higher terms [3,4].

Hereafter we propose for a wide and important class of potentials a simplified way of the improved calculation using only the same function which enters into the main approximation.

## 2. The semiclassical series expansion

Without loss of generality we can write the equation determining energy levels $\varepsilon_n$ in a potential well $V(x)$ in the following form:

$$\Phi(\varepsilon) = \frac{1}{\pi\beta}\int \sqrt{\varepsilon - V}\, dx = n + \frac{1}{2} + \delta \qquad (1)$$

with the unknown yet function $\delta$ which ensures the exact spectrum. Here $n = 0, 1, 2...$, the integral is taken between turning points and

$$\beta^2 = \hbar^2/2m \qquad (2)$$

The common popular semiclassical condition corresponds to $\delta \equiv 0$. It is known that such condition is exact only for few potentials. A power series expansion

$$\delta = \sum_{k=1}^{\infty} \delta_k$$
$$\delta_k = f_k(\varepsilon)\beta^{2k-1} \qquad (3)$$

Is known for $\delta$ but is not practically used since $\delta_k$ are very cumbersome for $k > 1$,

$$\delta_1 = \frac{\beta}{24\pi}\frac{\partial^2}{\partial \varepsilon^2}\int \frac{dx}{\sqrt{\varepsilon - V}}\left(\frac{dV}{dx}\right)^2. \qquad (4)$$

For all potentials which may be expressed by means of an auxiliary function $s(x)$ such that [4]

$$V(x) = A^2 s^2 + Bs + C$$

$$\sigma \equiv \frac{ds}{dx} = a_2 s^2 + a_1 s + a_0 \tag{5}$$

the value of $\delta_1$, calculated according to (4),

$$\delta_1 = \frac{\beta a_2}{8A} \tag{6}$$

and is invariant under the transformation $V \to V - C$, $s \to s + const$.

Hereafter we choose (with the same $a_2, A$)

$$V = A^2 s^2, \tag{7}$$

so that $x = 0$ is the point of a parabolic minimum for our potentials.

## 3. Calculation of $\delta_1$ by means of $\Phi(\varepsilon)$

Substituting into the integral (1)

$$dx = ds/\sigma \tag{8}$$

with $\sigma$ from (5) and $V$ (7), we obtain $\Phi(\varepsilon)$ in a special form depending on $A$ and an. In its turn we can regard each of these parameters as a definite function of all other parameters, $\varepsilon$ and $\Phi(\varepsilon)$.

In the present section we only treat even potentials so that $a_0 = 0$ and are interested in $a_2$.

We write at once $\delta_1$ (6), simply connected with $a_2$

$$\delta_1 = -\frac{1}{\Phi(\varepsilon)} \left( \frac{\varepsilon}{4\beta A \Phi(\varepsilon) a_0} - 1 \right). \tag{9}$$

For all potentials (5) $\delta_1$ does not depend on $\varepsilon$ though this parameter appears in (9). It is easy to prove that the bracket in (9) and thus $\delta_1$ are equal to zero if $a_2 = 0$.

For all such potentials is also valid at $x \to 0$ (10)
$$V(x) \to kx^2$$

Comparing (5) and (7) we get

$$a_0 = \frac{\sqrt{k}}{A} \qquad (11)$$

Thus we have obtained $\delta_1$ without differentiation and other operations of (4). It may be especially important for potentials given numerically. Besides, any differentiation makes worse analytic properties.

### 4. Adiabatic approximanion

Moreover, the quantization condition (1) is exact for all potentials satisfying to (5) if we put [4]

$$\delta = \frac{2\delta_1}{1+\sqrt{1+16\delta_1^2}}; \qquad (12)$$

It should be stressed that (5) is not a given ad hoc class but it embraces all the potentials used in handbooks as model and reference ones for the semiclassical approximation as well as for exact solutions. This class embraces the potentials related to the factorization method as well as to the supersymmetric theory with an additive parameter.

Comparing (3) and (12) we can also obtain all higher corrections as a series expansion in powers of $\delta_1$, e.g.

$$\delta_3 = -4\delta_1^3. \qquad (13)$$

Note once more that for all potentials (5) – and only for them - $\delta_1$ as well as $\delta$ do not depend on $\varepsilon$.

Suppose we study potentials which are near to the above class (5) so that $\delta_1$ calculated as earlier from (9) is a smooth function of $\varepsilon$. Then we can use previous expressions for $\delta_1(\varepsilon)$ and $\delta(\varepsilon)$ as an adiabatic approximation. A sufficient dimensionless condition for its validity will be one of the following inequalities:

$$d\delta_1/dn \ll 1, \qquad (14)$$

$$d\delta/dn \ll 1 \qquad (15)$$

It worth be mentioned that in an important extreme case of a very shallow potential well – in spite of (14), (15) - we become right values [4]

$$\delta_1 \to -\infty, \qquad \delta \to -½ \qquad (16)$$

Thus we have developed a simple way of the improved semiclassical calculation. It is exact for (5) and always in an extreme case (16). Since all or almost all interesting potentials are similar to (5), one may expect a good accuracy of this way practically in all cases.

We have supposed above that $a_1 = 0$. Tracing our calculations, we shall see that our way is always exact for (16) and approximately correct if

$$|a_1|^2 < |a_0 a_2|.$$

Various improvements and approximations proposed earlier [4,5] may be combined with our new one.

6.